\documentclass[aps,english,prd,showpacs,twocolumn,nofootinbib]{revtex4-1}
\usepackage{amsmath,amssymb}
\usepackage[usenames,dvipsnames]{xcolor}
\usepackage[latin9]{inputenc}
\usepackage{graphicx}
\usepackage{esint}
\usepackage{comment}
\usepackage{babel}
\usepackage{tensor}
\usepackage{ulem}
\usepackage{mathrsfs}
\usepackage{amsthm}
\usepackage{bigints}
\everymath{\displaystyle}
\usepackage{hyperref}
\makeatletter

\newcommand{\bd}{{\mathbf{d}}}

\newcommand{\bet}{{\boldsymbol{\eta}}}
\newcommand{\sm}{\scriptscriptstyle}

\newcommand{\non}{\nonumber\\}
\newcommand{\bLambda}{{\mathbf{\Lambda}}}
\newcommand{\bbeta}{{\boldsymbol{\beta}}}
\newcommand{\balpha}{{\boldsymbol{\alpha}}}
\usepackage{hyperref}
\@ifundefined{textcolor}{}
{%
 \definecolor{BLACK}{gray}{0}
 \definecolor{WHITE}{gray}{1}
 \definecolor{RED}{rgb}{1,0,0}
 \definecolor{GREEN}{rgb}{0,1,0}
\definecolor{dgreen}{rgb}{.1,.6,.1}
\definecolor{BLUE}{rgb}{0,0,1}
 \definecolor{CYAN}{cmyk}{1,0,0,0}
 \definecolor{MAGENTA}{cmyk}{0,1,0,0}
 \definecolor{YELLOW}{cmyk}{0,0,1,0}
 \definecolor{aw}{rgb}{0.2,0.5,0.75}
  }
  \definecolor{MyDarkRed}{rgb}{0.71,0.14,0.07}
\definecolor{MyDarkBlue}{rgb}{0.1,0,0.7}
\definecolor{MyDarkGreen}{rgb}{0.11,0.64,0.22}
 
 \hypersetup{
    colorlinks,%
    citecolor=blue,%
    filecolor=blue,%
    linkcolor=magenta,%
    urlcolor=blue
}

\makeatletter

\newcommand{\bg}{{\mathbf{g}}}

\newcommand{\bX}{{\mathbf{X}}}

\newcommand{\CP}{{\mathscr P}}
\newcommand{\SR}{{\mathscr R}}
\newcommand{\ST}{{\mathscr T}}

\newcommand{\CR}{{\cal R}}

\newcommand{\FP}{{\frak P}}

\newcommand{\initial}[1]{{#1_{\rm \bf i}}}

\def\t#1#2#3{#1^{\if#2- \else #2 \fi}_{\if#2- \else \:\: \fi #3}}
\def\tinv#1#2#3{#1^{\if#2- \else \:\: #2 \fi}_{\if#2- \else \fi #3}}
\def\tdot#1#2#3{\dot{#1}^{\if#2- \else #2 \fi}_{\if#2- \else \:\: \fi #3}}
\def\tddot#1#2#3{\ddot{#1}^{\if#2- \else #2 \fi}_{\if#2- \else \:\: \fi #3}}
\def\ttilde#1#2#3{\tilde{#1}^{\if#2- \else #2 \fi}_{\if#2- \else \:\: \fi #3}}

\usepackage{ifthen}
\newboolean{mag_part}
\setboolean{mag_part}{false} 
\newcommand{\magp}{\ifthenelse{\boolean{mag_part}}}
\newboolean{newton_com}
\setboolean{newton_com}{false} 
\newcommand{\newc}{\ifthenelse{\boolean{newton_com}}}

\newboolean{notusefull}
\setboolean{notusefull}{false} 
\newcommand{\nusefull}{\ifthenelse{\boolean{notusefull}}}

\newboolean{thirdorder}
\setboolean{thirdorder}{false} 
\newcommand{\third}{\ifthenelse{\boolean{thirdorder}}}

\newboolean{RZA}
\setboolean{RZA}{false} 
\newcommand{\rza}{\ifthenelse{\boolean{RZA}}}

\newcommand\eprintarXiv[1]{\href{http://arXiv.org/abs/#1}{#1}}

\makeatother

\begin{document}

\title{Lagrangian theory of structure formation in relativistic cosmology. \\IV.
Lagrangian approach to gravitational waves}

\author{Fosca Al Roumi$^{1}$, Thomas Buchert$^{1,*}$, and Alexander Wiegand$^{2,3,4}$}

\affiliation{$^1$Univ Lyon, Ens de Lyon, Univ Lyon1, CNRS, Centre de Recherche Astrophysique de Lyon UMR5574, F--69007, Lyon, France}

\affiliation{$^{2}$Max--Planck--Institut f\"ur Gravitationsphysik, Albert--Einstein--Institut,
 Am M\"uhlenberg 1, D--14476 Potsdam, Germany}
 
\affiliation{$^{3}$Harvard--Smithsonian Center for Astrophysics, 
60 Garden St., Cambridge MA 02138, U.S.A.}

\affiliation{$^{4}$Max--Planck--Institut f\"ur Astrophysik,
Karl--Schwarzschild--Str.~1, D--85741 Garching, Germany}

\affiliation{$^{*}$corresponding author: buchert@ens-lyon.fr}


 
\begin{abstract}
The relativistic generalization of the Newtonian Lagrangian perturbation theory is investigated. In previous works, the perturbation and solution schemes that are generated by the spatially projected gravitoelectric part of the Weyl tensor were given to any order of the perturbations, together with extensions and applications for accessing the nonperturbative regime. We here discuss more in detail the general first--order scheme within the Cartan formalism including and concentrating on the gravitational wave propagation in matter. We provide master equations for all parts of Lagrangian--linearized perturbations propagating in the perturbed spacetime, and we outline the solution procedure that allows one to find general solutions. Particular emphasis is given to global properties of the Lagrangian perturbation fields by employing results of Hodge--de Rham theory. We here discuss how the Hodge decomposition relates to the standard scalar--vector--tensor decomposition. Finally, we demonstrate that we obtain the known linear perturbation solutions of the standard relativistic perturbation scheme by performing two steps: first, by restricting our solutions to perturbations that propagate on a flat unperturbed background spacetime and, second, by transforming to Eulerian background coordinates with truncation of nonlinear terms. 
\end{abstract}

\maketitle

\section{Introduction}
\label{intro}

\vspace{-2pt}

In this series of papers we generalize the Lagrangian perturbation theory to relativistic cosmology by employing spatial Cartan coframes as the only perturbation variable. In \cite{rza1} we investigated $3+1$ first--order solutions for the trace and antisymmetric parts of perturbations of a dust fluid, and we proposed an extrapolation  in the spirit of Zel'dovich's approximation \cite{zeldovich,buchert89,bildhaueretal}.\footnote{In Newtonian cosmology in its Lagrangian form \cite{buchert:varenna,ehlersbuchert} this extrapolation is obtained by employing general functional expressions of the perturbed variable (see also the original proposal \cite{Kasai95} and the discussions in \cite{salopek} and \cite{rampf:zeldovich}; recall that a relativistic generalization of Zel'dovich's approximation can only be identified with a Lagrangian perturbation solution, if the full system of equations can be expressed in terms of a single variable \cite{rza1}).} In \cite{rza2} the resulting nonperturbative scheme of structure formation was then applied to quantify average properties of inhomogeneous cosmologies in relation to the Dark Energy and Dark Matter problems. In \cite{rza3} we gave the general perturbation and solution schemes at any order of the perturbations that cover the full hierarchy of the Newtonian Lagrangian perturbation and solution schemes. 

In this paper we proceed with a detailed discussion of the general first--order scheme concentrating on the trace--free symmetric parts of the perturbations and, thus, focusing on gravitational waves propagating in continuous matter.
The recent detections of gravitational waves by the LIGO/VIRGO teams \cite{ligovirgo} has heated up this research subject, and we here provide further steps toward a nonlinear comprehension of the theory of gravit\-ational waves; for a historical account and references see \cite{history}.  As in previous work we restrict our attention to irrotational dust continua for simplicity. The generalization to more general matter models and general foliations of spacetime is scheduled. 

\vspace{-2pt}

We recall the conceptual differences of our framework compared with the standard approach in relativistic cosmology (for a list of key--references on the standard approach as well as work that relates to our approach see \cite{rza3}). We choose the  
(spatially diffeomorphism invariant) Cartan coframes as a single perturbation variable, not the metric; we seek to describe the evolution of the perturbation fields intrinsically, thus operating on the physical Riemannian mani\-fold, not on a global background mani\-fold. As a consequence, perturbations cannot be expressed in terms of inertial (Eulerian) coordinates of the background, but in terms of local coordinates in the tangent spaces at each point of the physical manifold. These local coordinates take the role of Lagrangian coordinates in a flow--orthogonal (comoving) foliation of spacetime, adopted in this paper. With this approach we expect to capture nonlinear properties of gravitational waves, since a perturbative solution of these latter propagate in the perturbed spacetime, not in the background spacetime. This holds true even for linearized solutions for the Cartan coframes that we will consider throughout this paper. As for the gravitoelectric part of the relativistic solutions, studied in \cite{rza3}, we here will also demonstrate how the solutions of the gravitomagnetic part of the standard perturbation approach are recovered. The reader may consult the introduction to \cite{rza3} for further explanation of this approach.

This paper is structured as follows.
Section~\ref{sec:firstorder} recalls the basic system of equations, and provides the general perturbation scheme up to the first order; we rewrite this scheme in a couple of steps by employing decompositions in terms of symmetry and a split into gravitoelectric and gravitomagnetic parts; we then summarize the equations for each part in terms of master equations. Section~\ref{sec:hodge} provides an alternative global view on the perturbation scheme by employing Hodge--de Rham theory. 
Section~\ref{standard} performs a detailed comparison with the standard first--order approach to gravitational waves. Section~\ref{sec:conclusion} discusses the results with an outlook. In Appendix~\ref{sec:appA} we provide the Lagrangian propagation equations for the gravitoelectric and gravitomagnetic parts of the Weyl tensor.

\section{Foundations and first--order perturbation scheme}
\label{sec:firstorder}

First, we will provide a brief recapitulation of the basic system of equations and the perturbation ansatz.

\subsection{Foundations}

As in previous work, we employ a flow--orthogonal spacetime foliation admitting the following bilinear metric forms:
\begin{equation}
{}^{(4)}\bg = - \bd t \otimes \bd t + {}^{(3)} \bg \;, \text{ with }\; {}^{(3)} \bg = g_{ij} \,\bd X^i \otimes \bd X^j \;,
\end{equation}
where we denote by $X^i$ Gaussian normal (or Lagrangian) coordinates.
The spatial metric form is decomposed into Cartan coframes,
\begin{equation}
{}^{(3)}\bg = G_{ab} {\bet}^a \otimes {\bet}^b  \;\;,
\end{equation}
where $G_{ab} (\bX)$ is Gram's matrix that encodes the initial metric perturbations (see Equation~\eqref{metric} below).
The dynamical variable is composed of a $t-$parametrization of three spatial one--form fields.\footnote{Note that in place of orthonormal coframes we employ more general \textsl{adapted coframes}, see \cite{rza2,rza3} for more details.
Restricting the orthonormal spatial coframes ${\bet^a_{\rm on}}$ to exact forms ${\bet^a_{\rm on}} = \bd f^a$ implies for the $3-$metric, spatial diffeomorphism equivalence with a Euclidean space. This statement is obvious for
orthonormal coframes, but it also remains true for our \textsl{adapted coframes} and the requirement of being exact, as long as we assume exactness for the orthonormal coframes: the reader may consult Appendix A in \cite{rza3} for the proof of equivalence of this property for both types of coframes. Requiring exactness of the coframes also implies that the counterindices (denoted by $a,b,c \cdots$) can then be used as coordinate indices (denoted by $i,j,k \cdots$); the functions $f^a$ then define coordinates $x^i = f^{a\rightarrow i}$.}

The system of $3+1$ coefficient equations for the Einstein equations of an irrotational dust model, projected 
on the exact cotangent basis $\bd X^i$ reads:\footnote{We here only give the equations in coordinate components; for their representation in terms of differential forms the reader is directed to \cite{rza1,rza3}.} 
\begin{align}
\label{form_symcoeff}&G_{ab} \,\dot{\eta}^a_{[i} \eta^b_{\ j]} = 0 \;; \\
\label{form_eomcoeff}&\frac{1}{2 J} \epsilon_{abc} \epsilon^{ikl} \left( \dot{\eta}^a_{\ j} \eta^b_{\ k} \eta^c_{\ l} \right) \dot{} = -\CR^i_{\ j} + \left( 4 \pi G \varrho + \Lambda \right) \delta^i_{\ j}\;; \\
\label{form_hamiltoncoeff}&\frac{1}{2J}\epsilon_{abc} \epsilon^{mjk} \dot{\eta}^a_{\ m} \dot{\eta}^b_{\ j} \eta^c_{\ k} = - \frac{\CR}{2}+ \left( 8\pi G \varrho +  \Lambda  \right) \;; \\
\label{form_momcoeff}&\left(\tfrac{1}{J}\epsilon_{abc} \epsilon^{ikl} \dot{\eta}^a_{\ j} \eta^b_{\ k} \eta^c_{\ l} \right)_{||i} = \left(\tfrac{1}{J}\epsilon_{abc} \epsilon^{ikl} \dot{\eta}^a_{\ i} \eta^b_{\ k} \eta^c_{\ l} \right)_{|j} \;;
\end{align}
an overdot denotes a partial derivative with respect to the coordinate time (being equivalent to the covariant time--derivative in the present setting); a double vertical slash denotes the covariant spatial derivative with respect to the $3$--metric, the connection is assumed to be Levi--Civit\`a; the dust density is given by $\varrho = \initial{\varrho} J^{-1}$, $J \ge 0$, where the index ${}_\mathbf{i}$ marks initial data; $J$ defines the coefficient function of the $3-$volume form, $J \equiv \sqrt{g} / \sqrt{G}$, with
$\sqrt{g} \ \bd^3 X$ the $3-$volume form on the exact basis, $g: = \det (g_{ij}(\mathbf{X},t))$ and $G: = \det (G_{ij})= \det (g_{ij}(\mathbf{X},\initial t))$; $\CR_{ij}$ denote the coefficients of the spatial Ricci tensor, with $\CR$ its trace; the constant $\Lambda$ is the cosmological constant.

The set of equations $\lbrace$\eqref{form_symcoeff}--\eqref{form_momcoeff}$\rbrace$ is composed of $13$ equations, where the first $9$ equations 
are the needed evolution equations for the $9$ coefficient functions of the $3$ Cartan coframe fields, while the remaining $4$ equations 
originate from the energy and momentum constraints.

A key--insight underlying previous work is the possibility of paraphrasing formalisms and results of the Newtonian description in Lagrangian form \cite{buchertgoetz,buchert:varenna,ehlersbuchert} that allowed construction of a part of the relativistic perturbation solutions in \cite{rza3}.\footnote{This part, however,
completely covers all Newtonian solutions at any order, whose explicit expressions in this representation were given up to the fourth order \cite{buchert92,buchert93L,buchertehlers,buchert94,rampfbuchert}.} This insight is based on the formal analogy between the Lagrange--Newton system of equations and the following set of $4$ equations, in which the Ricci curvature is eliminated. It is composed of the first $3$ equations  \eqref{form_symcoeff}, and Raychaudhuri's equation (derived from the trace of the equation of motion \eqref{form_eomcoeff} combined with the energy constraint \eqref{form_hamiltoncoeff}):\footnote{Equations \eqref{form_symcoeff} imply $G_{ab} \,\ddot{\eta}^a_{[i} \eta^b_{\ j]} = 0 \ $.}
\begin{eqnarray}
G_{ab} \,\ddot{\eta}^a_{[i} \eta^b_{\ j]} = 0 \;;\nonumber \\
\label{LESgravitoelectric}\frac{1}{2J} \epsilon_{abc}\epsilon^{ik\ell} \ddot{\eta}^a_{\ i} \eta^b_{\ k} \eta^c_{\ \ell}   = \Lambda  - 4 \pi G {\varrho} \;.
\end{eqnarray}
For integrable coframe coefficients (or exact forms) the above system is closed and equivalent to the Lagrange--Newton system.
A short discussion of the interrelation between the Lagrange--Newton system and the above {\it gravitoelectric part} of the Lagrange--Einstein system can be found in \cite{mg14:fosca}.

\subsection{Perturbation ansatz}	

We follow the prescription of \cite{rza3} and decompose the coframes into a Friedmann--Lema\^\i tre--Robertson--Walker (FLRW) coframe and deviations thereof, the perturbations being {\it locally} developed to a given order $n$:
\begin{eqnarray}
 \bet^a = \eta^a_{\ i} \bd X^i = a(t) \left( \delta^a_{\ i} + \displaystyle\sum_n P^{a \sm(n)}_{\ i} \right) \bd X^i \;,
\end{eqnarray}
in the local basis $\bd X^i$. We use the normalization $a (t_\mathbf{i}) \equiv \initial a = 1$. The local metric coefficients are given by:\footnote{Recall that with this ansatz we choose to perturb a zero--curvature FLRW model, since it is possible to encode an initial first--order constant curvature in the coefficient functions $G_{ab}$ \cite{rza3}.}
\begin{eqnarray}
g_{ij} = G_{ab} \eta^{a}_{\ i} \eta^{b}_{\ j} \;\;;\;\; G_{ij} : = g_{ij} (t_\mathbf{i}) \;.
\label{metric}
\end{eqnarray}
with $G_{ij} = G_{ab} \,\eta^a_{\ i} (\initial t) \eta^b_{\ j} (\initial t) = G_{ab}\delta^a_{\ i}\delta^b_{\ j}$.

As the coframes are the only dynamical variables, we are entitled to employ functional definitions of all other variables such as the density, the metric or the curvature in terms of the deformation of the manifold at a given order. The strategy followed in previous work of this series was to inject the $n$--order coframes into these functional definitions without further truncation, furnishing nonperturbative approximations that may improve iteratively by going to higher order deformations. In this paper, however, we do not exploit this strategy and concentrate on strictly linearized equations.

The deformation coefficients $P^{a}_{\ i}$ only appear summed over the noncoordinate index in the equations, so we introduce the following tensor coefficients and their trace:
\begin{equation}
P^{i}_{\ j} \equiv \delta_{a}^{\ i} P^{a}_{\ j} \quad \textrm{and} \quad P \equiv {P}^{k}_{\ k} = {\delta}_{a}^{\ k} {P}^{a}_{\ k} \;,
\end{equation}
and we use this notation throughout this paper. 
Confining ourselves to first--order deviations, we have for the coefficients of the perturbation one--forms,
\begin{eqnarray}
\label{pert} \eta^a_{\ i} = a(t) \left( \delta^a_{\ i} +  P^{a}_{\ i} \right)  \;,
\label{firstorderdef}
\end{eqnarray}
and we have for the metric coefficients (\ref{metric}),\footnote{Note that only the zero--order metric tensor will appear in the  linearized equations.}
\begin{equation}
g_{ij}=a^2(t) \left( G_{ij} +2\,P_{(ij)} + P_{ai}P^a_{\:\:j}\right)\;;
\end{equation} 
we have defined:
\begin{equation}
P_{ij}:=G_{ai} P^a_{\:\:j}\:\:.
\end{equation}
For further details, especially concerning the initial data, the reader is directed to \cite{rza3}. 
However, we here briefly recall the notations for the general initial data set.
	
We have chosen to prescribe initial data (being considered, without loss of generality, first order) in terms of the $6$ one--form fields $\mathbf{U}^a = U^a_{\ i} \bd X^i$ and $\mathbf{W}^a = W^a_{\ i} \bd X^i$, being nonintegrable generalizations of the Newtonian peculiar--velocity and --acceleration gradients, and the initial values of the 
perturbations \lbrack$P^{a}_{\ i} (t_\mathbf{i}) =: \CP^a_{\ i}$\rbrack $\ $ together with their time--derivatives:\footnote{
$\dot{\CP}^{a}_{\ i}$ or $\ddot{\CP}^{a}_{\ i}$ is a shorthand of ${\dot P}^{a}_{\ i} (t_\mathbf{i})$ or ${\ddot P}^{a}_{\ i} (t_\mathbf{i})$, respectively.}
\begin{align}
         \begin{cases}
     \label{initialP} 
\CP^a_{\ i} = 0 \;; \\
\dot{\CP}^a_{\ i} = U^a_{\ i} \;,\; U_{[ij]} = 0 \;;\\ 
\ddot{\CP}^a_{\ i} = W^a_{\ i} - 2 \initial H U^a_{\ i}\;,\; W_{[ij]} = 0 \;,
         \end{cases}
\end{align}
with the Hubble function $H = \dot a / a$ at initial time, $\initial H = \dot {\initial a}$ (we recall the normalization of the scale factor to be $\initial a = 1$).
We use the abbreviations $\delta^k_{\ a } U^a_{\ k} = :U$, $\delta^k_{\ a } W^a_{\ k} = :W$ for the trace expressions.
The energy and momentum constraints have to be obeyed for admissible initial data:
\begin{align}
         \begin{cases}
\label{initialdata3}
    		\initial H U =-\frac{\SR}{4} - W \;; \\[.2cm]
	    	\left(U_{\; j}^{a}\delta_{a}^{\; i}\right)_{||i}=\left(U_{\; i}^{a}\delta_{a}^{\; i}\right)_{|j} \;,		
	\end{cases}
\end{align}
with the initial Ricci curvature as found from the equation of motion \eqref{form_eomcoeff},
\begin{equation}
\label{kinematicRicci}
 \CR_{ij}^{\sm(1)}(t_\mathbf{i}) = :\SR_{ij} = - (W_{ij} +\initial H U_{ij} ) - \delta_{ij} (W +\initial H U)\:\:.
\end{equation}
These initial data determine the problem completely: in our $3+1$ split, the system of equations of motion for the coframes 
$\lbrace$\eqref{form_symcoeff}--\eqref{form_momcoeff}$\rbrace$ is composed of the $6$ second--order differential equations \eqref{form_eomcoeff}. As $U_{ij}$ and $W_{ij}$ are symmetric $3-$matrices (originating from the equations \eqref{form_symcoeff}), they provide the $12$ initial conditions needed for the first and second derivatives of the coframes. For both, the inital data and the coframes there are $4$ constraints.

		\subsection{First--order equations}

The first--order Lagrange--Einstein system (the index $(1)$ is omitted for the perturbation fields but kept for the Ricci curvature) read:
\begin{eqnarray}
\label{symmetry} \dot{P}_{[ij]} = U_{[ij]} a^{-2} = 0 \;; \\
\label{full} \ddot{P}_{ij} + 3 H \dot{P}_{ij} = - a^{-2} \left( \CR^{\sm(1)}_{ij} - \frac{\CR^{\sm(1)}}{4} a^2 \delta_{ij} \right) \;;\\
\label{hamilton} H \dot{P} + 4\pi G \varrho_{H \rm \bf i} a^{-3} P = - \frac{\CR^{\sm(1)}}{4} - a^{-3} W \;; \\
\label{momentum} \dot{P}^i_{\ [i|j]} = 0 \;.
\end{eqnarray}
A more transparent representation of these equations decomposes the deformation fields into trace, trace--free symmetric and antisymmetric parts, 
\begin{eqnarray}
P_{ij} = P_{(ij)} + P_{[ij]} =  \frac{1}{3} P \delta_{ij} + \Pi_{ij} + \FP_{ij} \;,
\label{eq:decomp}
\end{eqnarray}
where we defined $\Pi_{ij} := P_{(ij)} - {1}/{3} \ P \delta_{ij}$, $\FP_{ij} : = P_{[ij]}$, and we introduce the trace--free symmetric part of the Ricci tensor, $\tau^{\sm(1)}_{ij} := \CR^{\sm(1)}_{ij} - {1}/{3} \ \CR^{\sm(1)}\delta_{ij}$. 
The first--order system for the deformation coefficients now reads:
\begin{eqnarray}
\label{symcondition_order1} \dot{\FP}_{ij} = U_{[ij]} a^{-2} \; 
&=& 0\;; \\
\label{trace_order1} \ddot{P} + 3 H \dot{P} \; &=& - \frac{\CR^{\sm(1)}}{4} \;; \\
\label{sympart_order1} \ddot{\Pi}_{ij} + 3 H \dot{\Pi}_{ij} \; &=& - a^{-2} \tau_{ij}^{\sm(1)} \;; \\
\label{hamilton_order1} H \dot{P} + 4\pi G \varrho_{H \rm \bf i} a^{-3} P \; &=& - \frac{\CR^{\sm(1)}}{4} - a^{-3} W\;; \\
\label{momentum_order1} \frac{1}{3} \dot{P}_{|j} - \frac{1}{2} \left(\dot{\Pi}^i_{\ j}\right)_{|i} \; &=& 0 \;.
\end{eqnarray}
In order to solve the {first--order} trace equation and the traceless symmetric equation, it is necessary to express the first--order scalar curvature and the traceless Ricci tensor ${}^{(1)}\tau_{ij}$. To do so, we inject the metric and its inverse, truncated to first--order:
\begin{eqnarray}
\label{metricorder1}g_{ij} &&= a^2 \left( \delta_{ij} + G^{\sm(1)}_{ij} + 2 P_{(ij)} \right)\;; \\
g^{ij} &&=  a^{-2} \left( \delta^{ij} - G^{ij \sm(1)} - 2 P^{(ij)} \right) \;,
\end{eqnarray}
into the definitions of the spatial Christoffel symbol and spatial Ricci tensor, also making use of the momentum constraints, to obtain:
\begin{eqnarray}
\Gamma^{k \ \sm(1)}_{\ ij} &&= \frac{1}{2} \delta^{kl} \left( G^{\sm(1)}_{li|j} + G^{\sm(1)}_{lj|i} - G^{\sm(1)}_{ij|l} \right) \\
&&+ \delta^{kl} \left( P_{(li)|j} + P_{(lj)|i} - P_{(ij)|l} \right)\;; \\
\CR_{ij}^{\sm(1)} &&= G_{i[k|j]}^{{\sm(1)} \ \ |k} + G^{k \sm(1)}_{\ [j|k]i} + P_{i[k|j]}^{\ \ \ \ |k} + P_{j[k|i]}^{\ \ \ \ |k} \;;\label{eq:riccifirst}\\
\CR^{\sm(1)} &&= 2a^{-2} G^{l {\sm(1)} \ |k}_{[k|l]} = :\frac{\SR}{a^2} \;. \label{eq:scalcurv}
\end{eqnarray}
Using the split into parts with different symmetries, we also express the curvature through the parts $P$ and $\Pi_{ij}$. A further manipulation, using again the momentum constraints, allows one to rewrite some terms in the curvature to obtain:
\begin{eqnarray}
\CR_{ij}^{\sm(1)} &&= \SR_{ij} + P_{|ij} - \frac{1}{3}  P^{|k}_{\ |k} \delta_{ij}- \Pi_{ij\ |k}^{\ \ |k}  \;;\label{fullRicci} \\
\CR^{\sm(1)} &&= a^{-2} \SR \;.
\label{fullscalarcurvature}
\end{eqnarray}
We can now write the first--order traceless part of the Ricci curvature tensor as:
\begin{eqnarray}
\tau_{ij}^{\sm(1)} &&=\ST_{ij} + P_{|ij} - \frac{1}{3} P^{|k}_{\ |k} \delta_{ij} - \Pi_{ij\ |k}^{\ \ |k} \;,
\label{tracelesscurv}
\end{eqnarray}
where we defined $\ST_{ij} : = \tau_{ij} (t_{i}): = \SR_{ij} - {1}/{3} \ \SR  \delta_{ij}$. 
These expressions for the intrinsic curvature show that first--order perturbations only generate traceless curvature in the course of time, while the trace is a conformal rescaling of the initial scalar curvature.

Knowing the {\it first--order} scalar curvature and the traceless Ricci tensor, we can inject them into the  first--order system for the deformation coefficients \eqref{symcondition_order1}--\eqref{momentum_order1}.
In this step we also perform the time--integration of the (trivial) antisymmetric part and of the momentum constraints, and apply the constraints on initial data \eqref{initialP} and \eqref{initialdata3}:\footnote{Note that there is no constant of integration appearing in Equation~\eqref{MOM}, because we chose our initial data in \eqref{initialP} such that $\boldsymbol{\CP}^{a}=0$.}
\begin{eqnarray}
\label{ASYM}\FP_{ij} = \FP_{ij} (\initial t) \;, && \; \FP_{ij}(\initial t) = 0 \;; \\
\label{TR}\ddot{P} + 3 H \dot{P} &&= - \frac{a^{-2} \SR}{4} \;;\\
\label{SYM}&& \hspace{-3cm} \ddot{\Pi}_{ij} + 3 H \dot{\Pi}_{ij} - a^{-2} \Pi_{ij\ |k}^{\ \ |k} = \non
&& \hspace{-2.3cm}
- a^{-2} \left( \ST_{ij} + P_{|ij} - \frac{1}{3} P^{|k}_{\ |k} \delta_{ij} \right) \;;\\
\label{HAM} H \dot{P} + 4 \pi G \varrho_{H \mathbf \mathrm i} a^{-3} P && = - \frac{a^{-2} \SR}{4} - a^{-3} W \;;\\
\label{MOM} \frac{2}{3} P_{|j} &&=\Pi^k_{\ j|k}  \;.
\end{eqnarray}
In the following subsection, we will separate the symmetric traceless deformation field $\Pi_{ij}$ into a part that is (through the momentum constraints) determined to have the same time--dependence as the trace $P$, and a part that encodes free gravitational waves. 

\subsection{First--order master equations}
\label{sec:master}

In this subsection, evolution and constraint equations will be combined into a set of equations for the symmetric parts of the first--order perturbation fields (the antisymmetric part is trivially solved). The result is a set of equations that we will call {\it master equations}.

\subsubsection{Master equations for the trace and traceless symmetric parts}

Inserting the energy constraint \eqref{HAM} into the evolution equation \eqref{TR} provides
the {\it master equation for the trace part} (Raychaudhuri's equation):
\begin{eqnarray}
\label{RAY} \ddot{P} + 2 H \dot{P} - 4 \pi G \varrho_{H \mathbf \mathrm i} a^{-3} P = a^{-3} W \;.
\end{eqnarray}
For the example of an Einstein--de Sitter universe model (EdS), the solution reads:
\begin{align}
P=&\frac{3}{5}\left[(U\initial t +\frac{3}{2}W\initial t^2)a\right.\left.-\left(U\initial t -W\initial t^2\right)a^{-\frac{3}{2}}-\frac{5}{2}W\initial t^2\right].
\label{eq:trGRfo}
\end{align}
In the same spirit, we combine the momentum constraints \eqref{MOM} with the traceless part of the evolution equation \eqref{SYM}. We eliminate the initial traceless curvature via the initial data set \eqref{kinematicRicci} and obtain the following {\it master equation for the traceless symmetric part}: 
\begin{equation}
\begin{split}
 \ddot{\Pi}_{ij}+3H\  \dot{\Pi}_{ij} - a^{-2}  \left(W_{ij}^{tl} +\initial H U_{ij}^{tl}  \right)\qquad\qquad \\
 = a^{-2} \left( {\Pi}_{ij} {}^{\vert k}_{\ \vert k}+\frac12 \delta_{ij} \  \Pi^k_{\ l \vert k}{}^{\vert l} - \frac32\  \Pi^k_{\ j\vert ki} \right)\:\:,
\label{Master}
\end{split}
\end{equation} 
with ${}^{tl}$ denoting the traceless part of the initial tensor fields.
We note that the trace and the traceless part of the perturbation fields are coupled via the momentum constraints \eqref{MOM}. Thus, their dynamics is not independent. Understanding how they are coupled is in the focus of the next subsection.

\subsubsection{The gravitoelectric and gravitomagnetic parts}

We now split the traceless symmetric part into a component that is coupled to the trace, ${}^E{\Pi}_{ij}$, and one that is decoupled from it, ${}^H{\Pi}_{ij}$. As was explicitly shown in \cite{rza3},  ${}^E{\Pi}_{ij}$ can be obtained from a generalization of the Newtonian trace solution. We will now show that it is separable in space and time and has the same time--dependence as the trace $P$. 
Furthermore, we will show that ${}^E{\Pi}_{ij}$ and ${}^H{\Pi}_{ij}$ are closely related to the gravitoelectric and gravitomagnetic parts of the spatially projected Weyl tensor (hence our index notations $E$ and $H$). 

From the momentum constraints \eqref{MOM} we infer that a decoupling of ${}^H{\Pi}_{ij}$ from $P$ implies that it must be divergencefree. We write \eqref{MOM} as
\begin{equation}
\label{MOMsplit}
\frac23 P_{\vert j} = {\Pi}^k_{\ j \vert k} = :\left( {}^E{\Pi}^k_{\ j \vert k}+{}^H{\Pi}^k_{\ j \vert k} \right) \:\:,
\end{equation}
which can then be divided into two separate constraints:
\begin{equation}
\frac23 P_{\vert j} ={}^E{\Pi}^k_{\ j \vert k} \quad ; \quad {}^H{\Pi}^k_{\ j \vert k} : =0\:\:.
\label{MCC}
\end{equation}
We split the initial data accordingly,
\begin{equation}
U_{ij}=:{^{E}U}_{ij}+^{H}U_{ij}\:\:;\:\:W_{ij} = : {^{E}W}_{ij}+^{H}W_{ij}\;.
\label{splitinitial}
\end{equation}
The initial traceless part of the Ricci curvature reads:
\begin{equation}
W_{ij}^{tl} +\initial H U_{ij}^{tl}=: \left({}^EW_{ij}^{tl} +\initial H {}^EU_{ij}^{tl}\right)+\left({}^HW_{ij}^{tl} +\initial H {}^HU_{ij}^{tl}\right)\:\:.
\end{equation}
The propagation of the constraint equations \eqref{MOMsplit} guarantees the preservation of the decomposition \eqref{MCC} in time.

This split can be carried through to the master equation \eqref{Master}. Inserting the superposition $\Pi_{ij} = {}^E{\Pi}_{ij}+{}^H{\Pi}_{ij}$, we first extract the divergencefree part,
which obeys the {\it master equation for the gravitomagnetic part}: 
\begin{equation}
{}^H\ddot{\Pi}_{ij}+3H \ {}^H\dot{\Pi}_{ij} - a^{-2}\ {}^H{\Pi}_{ij} {}^{\vert k}_{\ \vert k} =a^{-2}\left({}^HW_{ij}^{tl} +\initial H {}^HU_{ij}^{tl}  \right).
\label{mastermagnetic}
\end{equation}
This equation describes the propagation of gravitational waves and has the form of d'Alembert's equation with a damping term due to expansion---note the important difference to the standard perturbation approach, discussed in Section~\ref{standard}: this equation assumes this form in the {\it local} coordinates of the tangent spaces at each point of the manifold. 

The gravitoelectric part ${}^E{\Pi}_{ij}$ is the solution of: 
\begin{equation}
\begin{split}
{}^E\ddot{\Pi}_{ij}+3H\ {}^E\dot{\Pi}_{ij} - a^{-2}  \left({}^E W_{ij}^{tl} +\initial H {}^E U_{ij}^{tl}  \right)\qquad\qquad \\
= a^{-2} \left({}^E{\Pi}_{ij} {}^{\vert k}_{\ \vert k}+\frac12 \delta_{ij} \ {}^E\Pi^k_{\ l \vert k}{}^{\vert l} - \frac32\ {}^E\Pi^k_{\ j\vert ki} \right)\;.
\end{split}
\label{eqE}
\end{equation}
As was explicitly shown in \cite{rza3}, and as can be obtained from Equation~(\ref{MCC}), ${}^E{\Pi}_{ij}$ has the same time--dependence as the trace part $P$. It is therefore solution of the {\it master equation for the gravitoelectric part}, which is formally analogous to \eqref{TR} (using \eqref{initialdata3}):
 \begin{equation}
\label{masterelectric}
{}^E\ddot{\Pi}_{ij}+3H\ {}^E\dot{\Pi}_{ij} =a^{-2}  \left({}^EW_{ij}^{tl} +\initial H {}^EU_{ij}^{tl}  \right)\:.
\end{equation}
The equivalence of \eqref{eqE} and \eqref{masterelectric} implies with \eqref{MCC}: 
\begin{equation}
\Delta_{0}{}^E{\Pi}_{ij} =\mathcal{D}_{ij} P\:\:,
\label{linkmodes}
\end{equation}
where $\Delta_{0}$ denotes the ordinary Laplacian with respect to the local Lagrangian coordinates. The traceless spatial double derivative operator is defined as follows:
$$
\mathcal{D}_{ij} = \partial_i \partial_j - \frac13 G_{ij} \Delta_{0} \;. 
$$
(Here, $G_{ij} \approx \delta_{ij}$.)
In the next subsection we are going to discuss some important issues of interpretation of the solutions to the above equations. In particular, we recall the relation between the Newtonian Lagrangian perturbation theory and the gravitoelectric part ${}^E{\Pi}_{ij}$, we discuss the role of the gravitomagnetic part ${}^H{\Pi}_{ij}$ for the dynamical evolution of the traceless Ricci tensor, and we motivate the need for global considerations.  

\subsection{Discussion}
\label{discussionofmasterequations}

Interesting insights into the gravitoelectric part of the perturbation field, denoted by ${}^E{P}_{ij}$,  can be obtained from a formal analogy between the Lagrange--Newton--System and the {\it gravitoelectric part of the Einstein equations} ---see \cite{rza3}. This set of equations, which comprises \eqref{LESgravitoelectric} and which does not explicitly involve the spatial Ricci curvature tensor, can be expressed in terms of identities on the trace and the antisymmetric part of the gravitoelectric part of the Weyl tensor (\textit{cf.} Equations~(33) of \cite{rza3}). In the comoving synchronous foliation of spacetime considered here, this nonclosed set of equations becomes identical to the closed Lagrange--Newton--System after the execution of the `Minkowski Restriction'. This mathematical operation sends Cartan coframe fields $\boldsymbol{\eta}^a$ to exact forms $ \mathbf{d} f^a$ (the reader may recall its definition in \cite{rza3} and note that the perturbation field ${}^E{P}_{ij} = 1/3 P \delta_{ij} + {}^E{\Pi}_{ij} $ coincides with the gravitoelectric solution of \cite{rza3}). Reverting the `Minkowski Restriction' allows us to build the relativistic counterpart of Newtonian solutions to the Lagrangian perturbation theory: 
 \begin{equation}
 \label{MRsolution}
 \mathbf{d} f^i = a(t) \left(  \mathbf{d} X^i +  \mathbf{d} P^i \right) \:\:\longrightarrow\:\:  \boldsymbol{\eta}^a = a(t) \left(  \mathbf{d} X^a +    {}^E\mathbf{P}^a \right)\:. 
 \end{equation}  
Although ${}^E{P}_{ij}$ features a Newtonian time--behavior, the spatial coefficient functions are {\it a priori} nonintegrable in the relativistic case, so that its trace and its trace--free part separately generate curvature. 
However, we emphasize a subtle property of the gravitoelectric perturbation field:  by comparing Equations \eqref{eqE} and \eqref{masterelectric}, leading to the constraint \eqref{linkmodes}, with the defining equation for the Ricci tensor \eqref{fullRicci}, we observe that the gravitoelectric part ${}^E{\Pi}_{ij}$ exactly compensates the nonconstant part of the traceless Ricci curvature generated by $P$. Thus, only the gravitomagnetic perturbation ${}^H{\Pi}_{ij}$ encodes the dynamics of the Ricci curvature or its traceless part \eqref{tracelesscurv}, respectively.\footnote{
This holds up to a global contribution of the gravitomagnetic part that is generated by 
${}^E{\Pi}_{ij}$. To understand this latter remark we note that 
the gravitomagnetic part generated by ${}^E{P}_{ij}$ solves the following equation ({\it cf.} \cite{rza3}, App. C):
\begin{equation}
\Delta_0 H_{ij}({}^E{P}_{ij})  =0 \:\:,
\label{harmH}
\end{equation}
i.e., there is in general a harmonic global part of the solution, an issue that we will address in the next section.}

Furthermore, we point out that the master equations for both perturbation fields require global conditions to be solved, which leads us to consider the topology of the spatial hypersurfaces.
In the next section, we will therefore adopt a global description of the spatial sections. Results of Hodge--de Rham theory will be used to investigate the global properties of the Cartan coframes and to relate their Hodge decomposition to the local decomposition $P_{ij} ={}^E{P}_{ij}+ {}^H{\Pi}_{ij}$. Local considerations do not fully determine the physical content of ${}^{E} P_{ij}$ and  ${}^{H} \Pi_{ij}$, as the elliptic equations we encounter need topological boundary conditions to be solved.

\section{Global considerations: topology and Hodge--de Rham theory} 
\label{sec:hodge}

In the last section we considered the projection of the first--order equations on a local coordinate basis $\{\mathbf{d} X^i\}$, hence constraining the validity of the results to the regions of the manifold that can be covered by a single coordinate chart. 
Topological considerations enable us to specify the boundary conditions needed to globally solve the governing equations. Note that the need for topological considerations is not specific to the intrinsic description of perturbation fields defined on perturbed spatial sections;  boundary conditions have to be specified too in standard perturbation theory where the perturbation fields are described to evolve on flat space sections. In this latter case the topology usually adopted is that of a 3--torus (periodic boundary conditions).

In the following subsection we will consider spatial hypersurfaces that have a closed topology. Under this assumption it is possible to apply the Hodge decomposition of the Cartan coframes. As we will see, this decomposition will provide a new understanding of the global properties of perturbations.

\subsection{Hodge theorem and Thurston's geometrization program}
\label{hodge}

In order to set the stage for the Hodge decomposition of forms, some additional formalism is necessary and summarized in this subsection. For introductions into differential geometry and Hodge's theorem \cite{hodge}, the reader may refer to \cite{aubin,Lee,Nakahara:2003nw,Frankel}.

\subsubsection{Laplace--de Rham operator and the Weizenb\"ock formula for a $1-$form}

Let $\mathcal{M}$ be a closed Riemannian $n-$dimensional mani\-fold, and let us introduce two $p-$forms $\balpha$ and $\bbeta$. We define a global positive definite inner product,
\begin{equation}
(\balpha , \bbeta) = \int_\mathcal{M} \balpha \wedge * \bbeta\:\:,
\end{equation}
where the dual of the form $\boldsymbol{\beta}$ in an $n-$dimensional space is $*  \boldsymbol{\beta}$:
\begin{equation}
* \boldsymbol{\beta} = \frac{1}{(n-p)!} \varepsilon_{\nu_1 \cdots \nu_p \ \mu_1 \cdots \mu_{n-p}} \beta^{\vert \nu_1 \cdots \nu_p \vert } \boldsymbol{\eta}^{\mu_1}\wedge  \cdots \wedge \boldsymbol{\eta}^{\mu_{n-p}}\:\:;
\end{equation}
$\{{\boldsymbol{{\eta}}}^{i}\}$, $i=1 \cdots n$ are the Cartan coframes defined on an $n-$dimensional manifold $\mathcal{M}$.

The Laplace--de Rham operator $\mathbf{\Delta}^{\rm dR} $ is a mapping from $p-$forms to $p-$forms; it generalizes the simple Laplacian from flat to curved manifolds. It is defined by:
\begin{equation}
\mathbf{\Delta}^{\rm dR} :=  \bd \bd^* + \bd^*  \bd = \left( \bd+\bd^*   \right)^2\:\:, 
\end{equation} 
where the codifferential operator $\boldsymbol{\delta} = \bd^*$ acting on the $p-$form $\bbeta$  is defined, on closed Riemannian manifolds, as follows:
\begin{equation}
\bd^* \bbeta = (-1)^{n(p+1)+1} * \bd * \bbeta\:\:.
\end{equation}
According to the Weizenb\"{o}ck formula, $\mathbf{\Delta}^{\rm dR} $  applied to three $1-$form fields $\boldsymbol{\eta}^{a}$, which is the case of interest for the Cartan coframes, yields in components, projected onto the exact basis $\lbrace\bd X^i \rbrace$:
\begin{equation}
\label{laplacederham}
\left( \mathbf{\Delta}^{\rm dR}\ \boldsymbol{\eta}^{a} \right)_i =  \left(\mathbf{\Delta}\boldsymbol{\eta}^{a}\right)_i \; + \; \eta^a_{\ k}\mathcal{R}^{k}_{\ i}\:\:,
\end{equation}
where $\mathcal{R}^{k}_{\ i}$ are the  $3-$Ricci curvature tensor components. The first term features the `rough Laplacian',\footnote{We here denote by $\mathbf{\Delta}$ the `rough Laplacian' (as the negative Bochner Laplacian) whose action on the coframe components can be written $-\boldsymbol{\eta}^{a}_{\ i}{}^{\parallel k}_{\ \ \parallel k}$, where $ {}_{\parallel}$ denotes the Lagrangian covariant derivative with respect to the spatial metric.} which implicitly depends on the geometry through the covariant derivatives, and the second term explicitly takes into account the local geometry through the Ricci curvature.

\subsubsection{The Hodge theorem}

The Hodge theorem \cite{hodge} asserts that on a closed, oriented and smooth Riemannian $n-$dimensional manifold, equipped with a smooth metric, the vector space of {\it harmonic $p-$forms}\footnote{Since
$
(\mathbf{\Delta} \balpha^p ,\balpha^p  ) = \parallel \bd^*\balpha  \parallel^2 + \parallel \bd \balpha  \parallel^2\  \geq 0\:,
\mathbf{\Delta}  \balpha^p = \mathbf{0}\:\: \mbox{if and only if} \:\:  \bd^*\balpha= \mathbf{0}\:\: \mbox{and} \:\:\bd \balpha  = \mathbf{0}\:.$
Harmonic forms are both closed and coclosed.} $\boldsymbol{\mathcal{H}}^p$  is of finite dimension. The generalized Poisson equation for $p-$forms $\balpha^p$ and $\boldsymbol{\rho}^p$ on a curved manifold,
\begin{equation}
\mathbf{\Delta}^{\rm dR}\ \balpha^p = \boldsymbol{\rho}^p\:\:,
\end{equation}
has a solution $\balpha^p$ if and only if $\boldsymbol{\rho}^p$ is orthogonal to $\boldsymbol{\mathcal{H}}^p$.  

For any $p-$form $\boldsymbol{\gamma}^p$ there exists a $p-$form $\boldsymbol{\theta}^p$ such that $\boldsymbol{\gamma}^p$ can be Hodge--decomposed as follows:
\begin{equation}
\boldsymbol{\gamma}^p = \mathbf{d}\balpha^{p-1} + \bd^* \bbeta^{p+1} + \mathbf{h}^p \;\;,
\label{hodge}
\end{equation}
where $  \balpha^{p-1} =\bd^* \boldsymbol{\theta}^p $,  $\bbeta^{p+1} = \bd \boldsymbol{\theta}^p $, and $ \mathbf{h}^p $ is harmonic. 
In other words, the vector space of $p-$forms can be decomposed as:
 \begin{equation}
\bLambda^{p}= \mathbf{d}\bLambda^{p-1} \bigoplus^\perp \bd^* \bLambda^{p+1}  \bigoplus^\perp \boldsymbol{\mathcal{H}}^p \;\;,
\end{equation}
where $\mathbf{d}\bLambda^{p-1}$, $\bd^* \bLambda^{p+1}$ and $\boldsymbol{\mathcal{H}}^p$ are the $p-$dimensional exact, coexact and harmonic vector spaces.

In view of what follows, we remark that the manifold need not to be simply--connected for employing the Hodge decomposition.

\subsubsection{Harmonic forms, and the geometry and topology of the Universe}
\label{harmonic}

In order to apply the Hodge decomposition \eqref{hodge}, we first have to address the question of the harmonic part that, if nonvanishing, would make a decomposition into exact and coexact forms nonunique. In general relativity, for regular Einstein flows, it is generally assumed that the topology of spacetime is given by the topology of the initial (Cauchy) hypersurface $\Sigma$, being conserved in time (hyperbolicity of Einstein's equations), $\Sigma \times {\mathbb R}$. Hence, we have to specify the topology of the Cauchy surface.

If this initial hypersurface is flat, then there is a simple reasoning to remove the harmonic part \cite{buchertehlers}: in the case of a flat space, the harmonic part is a vector field that obeys the simple Laplace equation, $\Delta h^a =0$. Solutions of the Poisson equation on a flat space are unique, if we impose periodic boundary conditions, i.e, we impose a $3-$torus topology to close the spatial section. Harmonic vector fields are then spatially constant, and this constant can be set to zero without loss of generality by exploiting the translational invariance of the equations on a flat space (e.g., Newton's equations have the same physical content if we add to the velocity model a spatially constant, eventually time--dependent vector field at each point).

For curved manifolds the vanishing of harmonic forms cannot be based on simple arguments. 
There are theorems (see, e.g., \cite{aubin}) stating that, if the oriented closed Riemannian manifold $\mathcal{M}$ has everywhere a positive Ricci tensor, then a harmonic $1-$form will vanish identically. This situation is, however, nongeneric in cosmology, since the Ricci curvature will alternate between positive and negative values.
If generic inhomogeneities are present, one
may (in line with Thurston's geometrization program \cite{thurston}, see below, and also \cite{LachiezeRey:1995kj}) conceive the idea of using regionally different backgrounds in our perturbative analysis. Here, ``regional background'' refers to length scales set by an average over
a given region containing inhomogeneities governing the background, see \cite{regionalbackground}. Such a formulation would allow us to construct a link to 
Thurston's geometrization program that asserts that every closed $3$--dimensional manifold can be decomposed into a connected sum of $3$--manifolds modeled after the $8$ model geometries listed by Thurston. 
Thurston's program was proven by Perelman in 2003 using Ricci flow with surgery \cite{perelman}.
For its realization one has to devise a way to match the different ``homogeneity patches'', a possibly difficult task.
In such a case we would have different situations for the harmonic forms (depending on the  regional background).
Perelman's work also implies a proof of Poincar\'e's conjecture: Thurston's geometrization program also asserts that  closed {\it and} simply--connected $3$--manifolds are homeomorphic to the hypersphere $\mathbb{S}^3$.
Since $\mathbb{S}^3$ admits the round metric (which has positive Ricci curvature), it can be shown that the only harmonic $1-$form is the null--form.\footnote{This is a direct consequence of the theorem by 
de Rham \cite{deRham,Bredon}:
If $v$ is a harmonic vector field on a compact Riemannian manifold and if the Ricci curvature $Ric(v,v)\geq 0$, then $v$ is parallel and $Ric(v,v)=0$. This in turn implies that there exists no nonzero harmonic field on a compact Riemannian manifold of positive Ricci curvature. Hence, since $\mathbb{S}^3$ admits the round metric, the first de Rham cohomology group $\mathcal{H}^1_{dR}(\mathcal{M}) =0$ (Hodge theory indeed states that the space of the harmonic $1-$forms $ \mathcal{H}^1(\mathcal{M})$ is isomorphic to the first de Rham cohomology group $\mathcal{H}^1_{dR}(\mathcal{M})$) \cite{Voisin,hodge}.}

The assumption of simple--connectivity has obvious limitations. For instance, it is valid for a constant--curvature space with positive curvature, or for infinite models with constant curvature (and trivial fundamental groups), but already for a toroidal model with zero curvature or negative constant curvature spaces that are finite, it fails. An alternating and inhomogeneous curvature is the generic situation.
Notwithstanding, we henceforth assume an $\mathbb{S}^3$ topology for the spatial sections, such that the harmonic part vanishes. This assumption is to be considered as a representative example of space--forms for which the vanishing of the harmonic part can be rigorously proved. (Recall that the vanishing of the harmonic part on a curved manifold generalizes the assumption made in Newtonian cosmology or standard perturbation theory.)

\subsection{Hodge decomposition of the Cartan coframes}
\label{cartandecomposition}

As we have seen in the previous subsection, the Hodge theorem allows us to decompose the  Cartan coframes into exact, coexact and harmonic forms. In the case of a closed {\it and} simply--connected manifold, e.g., $\mathbb{S}^3$, the harmonic $1-$form in this decomposition is set to zero. Furthermore, since any closed $3-$dimensional manifold is {\it parallelizable}, the Cartan coframes are continuously defined on the manifold while the coordinate charts undergo singularities. This case allows us to extend the definition of the coframes globally and apply to them the Hodge decomposition.\footnote{The $\mathbb{S}^3$ topology is a known example in cosmology for finite universe models with constant positive curvature such as the classical Einstein or Eddington universe models. Speaking in terms of constant--curvature template geometries for the Universe, this assumption may be linked to observations (a marginally spherical model \cite{gausmann} has been favored by Cosmic Microwave Background measurements of a reported first Doppler peak that is shifted by a few percent toward larger angular scales with respect to the peak predicted by the standard Cold Dark Matter model \cite{spherical}). For the opposite case of a hyperbolic universe model with small negative curvature see \cite{aurich01},
and for recent constraints on the constant--curvature universe model from the \textit{Planck} mission, see \cite{planck}.}

With these assumptions, the three Cartan coframes, which are three $1-$forms, can be decomposed as follows:
\begin{equation}
\boldsymbol{\eta}^a =  \mathbf{d}\alpha^a+  \bd^* \bbeta^a =  \boldsymbol{\Delta}^{dR} \ \boldsymbol{\gamma}^a \;\;,
\label{eq:hodgeeta} 
\end{equation}
where $\alpha^a$ are three scalars, $ \bbeta^a $ are three $2-$forms, and $\boldsymbol{\gamma}^a $ are three $1-$forms.

For later reference we project Equation~(\ref{eq:hodgeeta}) onto our local Lagrangian basis $\{\mathbf{d} X^i\}$,
$g:= \det (g_{ij})$:
\begin{eqnarray}
&\boldsymbol{\eta}^a =\alpha_{\; |j}^{a}\,\mathbf{d}X^{j}\nonumber\\
&+\left(\beta_{km}^{a}\epsilon^{pkm}g_{rp}\sqrt{g}\right)_{|l}\ g_{jn}\epsilon^{nrl}\sqrt{g}\ \mathbf{d}X^{j}\;,\quad
\label{eq:etadecomp} 
\end{eqnarray}
where the $2-$form coefficient matrix $\beta_{\ km}^a$ occurs in the projection of the $2-$form $\bbeta^a$ as $\bbeta^a=\beta_{\ km}^a \ \mathbf{d}X^k\wedge \mathbf{d}X^m$.

In three dimensions, each antisymmetric coefficient matrix $\beta^a_{\ km}$ has only three independent components. Writing these components as $B_{\ v}^{a}$, we can choose to express $\beta_{\ km}^{a}$ by $\beta_{\ km}^{a}=1/ 2 g^{wv}\sqrt{g}\epsilon_{wkm}B_{\ v}^{a}\;$. With this substitution the coefficient form of $\boldsymbol{\eta}^a$ reads:
\begin{equation}
\boldsymbol{\eta}^a =\alpha_{\; | j}^{a}\,\mathbf{d}X^{j}+ B_{\ u | l}^{a}\sqrt{{\rm det}\left(g\right)}g_{kj}\epsilon^{ulk}\ \mathbf{d}X^{j} \;.
\label{eq:etadecomp} 
\end{equation}

\subsubsection{Consequences for the intrinsic spatial curvature}
\label{curvatureconsequences}

The Hodge decomposition bears interesting consequences in the context of our decomposition into gravito\-electric and gravitomagnetic parts. From \eqref{harmH} we conclude that, for topologies with a null--dimensional harmonic space of $1-$forms, the gravitomagnetic part generated by ${}^E \mathbf{P}^a$ is zero. 
For this we note that the structure coefficients associated with the Cartan coframes,
\begin{equation}
\label{structC}
\mathbf{d}\, \boldsymbol{\eta}^a = -\frac12 C^a_{\ bc}\, \boldsymbol{\eta}^b \wedge \boldsymbol{\eta}^c\:\:,
\end{equation} 
are related, at first--order, to the (spatially projected) gravitomagnetic part of the Weyl tensor as follows:
\begin{equation}
H^i_{\ j} = \frac12 \epsilon_j {}^{kl}\dot{C}^i_{\ lk}\:\:.
\label{structH}
\end{equation} 
Since the gravitoelectric perturbation ${}^E \mathbf{P}^a$ generates only a harmonic gravitomagnetic part (that vanishes according to our global assumption), and since from \eqref{structH} the structure coefficients calculated from ${}^E \mathbf{P}^a$  are constant in time and are initially zero ($\boldsymbol{\eta}^a (\initial t) = \mathbf{d} X^a$),  we then conclude from \eqref{structC} that ${}^E \mathbf{P}^a$ represents the integrable part $\mathbf{d}\alpha^a$ of the perturbation fields. This is in accord with our finding that the curvatures generated separately by the trace and the trace--free gravitoelectric parts exactly compensate each other, see subsection \ref{discussionofmasterequations}. Thus, this global consideration confirms that the only curvature--generating part of the perturbation fields is ${}^H \mathbf{\Pi}^a$ (at first order).

Note that the scalar part of the curvature \eqref{eq:scalcurv} is at first order not part of the perturbation fields, but included in the normalization of the coframes $G_{a b}$. This defines the curved reference space for our first--order perturbation fields. In \cite{rza3} we explicitly demonstrated that, with respect to this space, the time--evolution of the perturbations correspond to the Newtonian solutions. With the decomposition \eqref{eq:hodgeeta} this can now be made more rigorous: the exact part can, as in the Newtonian case, be absorbed into a redefinition of global coordinates, leaving us this time not with a reparametrization of flat space, but rather of the initially perturbed metric $G_{i j}=g_{i j}(\initial t)$. The additional component in the nonexact part, that had not been considered in \cite{rza3}, encodes the nonintegrable geometrical deviations from the zero--order solution, featuring gravitational waves.

\subsubsection{Obtaining the Hodge fields from the perturbed coframes}

The Hodge decomposition of the Cartan coframes \eqref{eq:hodgeeta} can equivalently be defined for the perturbation fields. In our case these are taken to be of first order. Thus, we face the decomposition (not changing the notation for $\alpha^a$ and $\bbeta^a$ for simplicity):
\begin{equation}
\mathbf{P}^a =  \mathbf{d}\alpha^a +  \bd^* \bbeta^a \;\;.
\end{equation}
We project these fields onto the local Lagrangian basis $\{\mathbf{d} X^i\}$, and using again the fact that we work at first order, we find:
\begin{equation}
P^a_{\ j} = \alpha^a_{\ \vert j} + 2 \beta_{\ kj|l}^{a} \delta^{lk}\:\:,
\end{equation}
where $\alpha^a$ are three scalar fields and $\beta_{\ kj}^{a}$ the three $2-$form coefficient matrices $\bbeta^a=\beta_{\ km}^a \ \mathbf{d}X^k\wedge \mathbf{d}X^m$. Note that we used the fact that we work at first order and so $P^a_{\ j}$, $\alpha^a$ and $\beta_{\ kj}^{a}$ are first--oder quantities and $g_{ij}\approx a^2 \delta_{ij}$. In three dimensions, each antisymmetric coefficient matrix $\beta^a_{\ km}$ has only three independent components. Writing these components as $B_{\ v}^{a}$, we can express the matrix by $\beta_{\ km}^{a}=1/ 2 g^{wv}\sqrt{g}\epsilon_{wkm}B_{\ v}^{a}$. With this substitution the coefficient form of $P^a_{\ j}$ reads:
\begin{equation}
P^a_{\ j} = \alpha^a_{\ \vert j} +\delta_{rj}\, \epsilon^{ulr}\,{\rm{B}}^{a}_{\ u \,\vert \; l}\:\:,
\label{order1}
\end{equation}
again at first order. Formally the coefficients $B_{\ v}^{a}$ are attributed to the $1-$form fields ${\mathcal{\mathbf{B}}}^a$ and related to $\bbeta^a$ through the duality of $2-$ and $1-$forms in three dimensions.

Since $ {}^E\mathbf{P}^a$ encodes the integrable part of the perturbations, $ {}^E\mathbf{P}^a = \mathbf{d}\alpha^a $, it is straightforwardly related to the scalar modes of the standard perturbation approach ({\it cf.} Section~\ref{compaSVT}).  The complementary part, $ {}^H\mathbf{\Pi}^a$, is equal to $\boldsymbol{\delta} \bbeta^a$.

From \eqref{order1} we conclude that the Hodge fields $\alpha_{i}$ and ${\rm{B}}_{ij}$ can be obtained from the perturbation fields by solving the following Poisson equations in local (Lagrangian) coordinates:
\begin{equation}
\Delta_0 \alpha_j =P_{ij}{}^{\vert i}\:\:,
\label{consistent}
\end{equation}
\begin{equation}
\Delta_0 {\rm{B}}_{ij} =\epsilon_{j}{}^{mr}  P_{ir\; \vert m}     \:\:,
\label{eqB}
\end{equation}
where we have chosen the gauge condition on ${\rm{B}}_{ij}$ to be ${\rm{B}}_{i}^{\ m}{}_{\vert m} = 0 $.\footnote{This gauge degree of freedom comes from the fact that the physical content of $\mathbf{B}^a$ is determined by its relation to $\mathbf{P}^a$, given by Equation (\ref{order1}). In other words, due to the 
Levi--Civit\`a symbol, the term ${\rm{B}}_{i}^{\ m}{}_{\vert m}$ does not intervene in Equation (\ref{order1}); we are free to choose its value. Whatever the value of ${\rm{B}}_{i}^{\ m}{}_{\vert m}$, it will equally contribute to $\mathbf{P}^a$.}

The spatial gravitomagnetic part of the Weyl tensor can be expressed in terms of the expansion tensor (\textit{cf.} Equation (70) of \cite{rza1}): 
\begin{equation}
    H^{i}_{\,j} = -  \frac{1}{J} \epsilon^{ikl} \Theta_{jk \parallel l}\:\:.
\end{equation}
At first order, this yields:
\begin{equation}
    H^{i}_{\,j} = -  \frac{1}{a} \epsilon^{ikl} \dot{P}_{jk \vert l} \quad \mbox{and} \quad H_{ij} = - {a}\, \epsilon_i^{\: kl} \dot{P}_{jk \vert l} \:\:.
\end{equation}
We conclude that the first--order gravitomagnetic part is given by
\begin{equation}
    H_{ij} =  - a \Delta_0 \dot{\rm{B}}_{ij}    \:\:.
\end{equation}

\section{Comparison with the standard perturbation scheme}
\label{standard}

In this section we will explain the link between our approach and the standard one that uses the Scalar--Vector--Tensor 
(S--V--T) decomposition of the perturbed metric. In order to do so, we compare our results to the ones obtained in the comoving synchronous gauge being the choice of coordinates that initially coincides with the Lagrangian coordinates in the $3+1$--foliation we consider.
(At places where we refer to explicit solutions, we will adopt an EdS background solution as an example, without further indication in the text.)

\subsection{S--V--T decomposition in the comoving synchronous gauge}
\label{compaSVT}

In most of the literature, the perturbative approach is applied to 
the metric instead of the coframes (as the more elementary variable). 
The synchronous line--element is decomposed as:
\begin{equation}
{\mathrm d}s^{2}=-{\mathrm d}t^{2}+a^{2}(t)\gamma_{ij}({\bf x},t){\mathrm d}x^{i}{\mathrm d}x^{j}\;.
\end{equation}
The first--order corrections to a flat FLRW background are then written as:
\begin{equation}
\gamma_{ij}=\delta_{ij}+\gamma_{{\scriptscriptstyle {\rm S}}\, ij}\;.
\end{equation}
In the standard perturbation framework, it is common
to split the perturbations into scalar, vector and tensor contributions, rather than to consider a split into gravitoelectric and gravitomagnetic parts. The perturbation to the background can be decomposed as: 
\begin{equation}
\gamma_{{\scriptscriptstyle {\rm S}}\, ij}=-2\phi_{{\scriptscriptstyle {\rm S}}}\delta_{ij}+\mathcal{D}_{ij}\chi_{{\scriptscriptstyle {\rm S}}}^{\|}+\partial_{i}\chi_{{\scriptscriptstyle {\rm S}}\, j}^{\bot}+\partial_{j}\chi_{{\scriptscriptstyle {\rm S}}\, i}^{\bot}+\chi_{ij}^{\top}\;,
\label{sptmetcorr}
\end{equation}
where we follow the notation of \cite{Matarrese98} and denote the traceless derivative by $\mathcal{D}_{ij}=\partial_{i}\partial_{j}-\tfrac{1}{3}\Delta\delta_{ij}$, where $\Delta$ denotes the ordinary flat--space Laplacian in Eulerian background coordinates,
and 
\begin{equation}
\partial^{i}\chi_{{\scriptscriptstyle {\rm S}}\, i}^{\bot}= 0 \;\;;\;\; \chi_{~~i}^{\top i}=0\;\;;\;\;\partial^{i}\chi_{ij}^{\top}=0\;.
\end{equation}
As we consider only first--order corrections to the metric in this subsection, $\phi_{{\scriptscriptstyle {\rm S}}}$, $\chi_{{\scriptscriptstyle {\rm S}}}^{\|}$, $\chi_{{\scriptscriptstyle {\rm S}}\, j}^{\bot}$ and $\chi_{ij}^{\top}$ are first--order quantities as well.

The decomposition \eqref{sptmetcorr} simplifies for irrotational fluids. In this case the vector perturbations represent gauge modes that can be set to zero \cite{Matarrese98}. This result is in accord with the simplification of our coframe decomposition \eqref{eq:decomp}. Note that Equation~\eqref{eq:decomp} does not contain any vector fields anyway, as we have a local coordinate system (in the tangent space at a point in the Riemannian manifold) in which vector fields simply are not present in a flow--orthogonal foliation. However, the part that corresponds to the vector perturbations, the antisymmetric tensor part, also vanishes for the irrotational case.

With the vectors removed, we only need the synchronous gauge expressions for $\phi_{{\scriptscriptstyle {\rm S}}}$,
$\mathcal{D}_{ij}\chi_{{\scriptscriptstyle {\rm S}}}^{\|}$ and $\chi_{ij}^{\top}$
as given in \cite{Matarrese98}.
Let us start with the tensor part. It satisfies the equation
\begin{equation}
{\ddot{\chi}_{ij}^{\top}}+3H{\dot{\chi}_{ij}^{\top}}-\frac{1}{a^2}\Delta\chi_{ij}^{\top}=0\;.\label{eq:GWtEvol}
\end{equation}
This equation corresponds to our 
Equation~\eqref{mastermagnetic} in the sense explained further below.
The solution of (\ref{eq:GWtEvol}) reads \cite{Matarrese98}: 
\begin{equation}
\chi_{ij}^{\top}({\bf x},t)=\frac{1}{(2\pi)^{3}}\int d^{3}{\bf k}\exp(i{\bf k}\cdot{\bf x})\chi_{\sigma}({\bf k},t)\epsilon_{ij}^{\sigma}(\hat{{\bf k}})\;,
\end{equation}
with $\hat{\bf k}$ a normalized $\bf k$--vector, and where $\epsilon_{ij}^{\sigma}(\hat{{\bf k}})$ introduces the polarization
tensor; $\sigma$ is ranging over the polarization components $+$ and $\times$.
The $\chi_{\sigma}({\bf k},t)$ are the corresponding amplitudes.
The time--evolution is obtained from (\ref{eq:GWtEvol}) and provides the solution in terms of the first spherical Bessel function,
$j_{1}\left(x\right)= \sin x / x^{2} - \cos x / x$:
\begin{equation}
\chi_{\sigma}({\bf k},t)=A(k)a_{\sigma}({\bf k})\left(\frac{3j_{1}\left(3k\initial{t}^{2/3}t^{1/3}\right)}{3k\initial{t}^{2/3}t^{1/3}}\right)\;.\label{eq:freegw-2}
\end{equation}
$a_{\sigma}({\bf k})$ denotes a zero
mean random variable, and $A(k)$ encodes the form of the spectrum of primordial gravitational waves.

Second, we turn to the scalar sector. The solutions given in \cite{Matarrese98}
are restricted to the growing mode only. In addition, the authors
use the residual gauge freedom of synchronous gauge to fix $\chi_{S}^{\|}$
such that $\Delta\chi_{{\scriptscriptstyle {\rm S}}{\rm\bf i}}^{\|}=-2\initial{\delta}$.
Then, their solutions for the scalar sector are given by
\begin{equation}
\mathcal{D}_{ij}\chi_{{\scriptscriptstyle {\rm S}}}^{\|}=-3\initial{t}^{2}\left(\frac{t}{\initial{t}}\right)^{\frac{2}{3}}\biggl( \partial_{ij}\varphi-\frac{1}{3}\delta_{ij}\Delta\varphi\biggr)\;,\label{eq:chis-2}
\end{equation}
where $\varphi$ is time--independent and defined by its relation to $\initial{\delta}$ in the
cosmological flat--space Poisson equation $\Delta\varphi({\bf x})=2 / (3\initial{t}^{2})\initial{\delta}({\bf x})$, and the solution for the second scalar mode:
\begin{equation}
\phi_{{\scriptscriptstyle {\rm S}}}({\bf x},t)=\frac{5}{3}\varphi({\bf x})+\frac{1}{2}\initial{t}^{2}\left(\frac{t}{\initial{t}}\right)^{\frac{2}{3}}\Delta\varphi({\bf x})\;.
\end{equation}

\subsection{Comparing the S--V--T decomposition to the intrinsic perturbation fields}

In our case, $P_{ij}$ as well as  the initial metric $G_{ij}$ appear in the decomposition of the metric perturbation that we can formally compare:
\[
\gamma_{{\scriptscriptstyle {\rm S}}\, ij} \leftrightarrow G_{ij}+2P_{\left(ij\right)}\;.
\]
The relation between the two sets of perturbation fields can be formally obtained by a replacement of the dependent variables:
\begin{eqnarray*}
\phi_{{\scriptscriptstyle {\rm S}}} &  \leftrightarrow  & -\frac{1}{3}\left(P+\frac{1}{2}G \right)\;\;\\
\frac{1}{2}\left(\mathcal{D}_{ij}\chi_{{\scriptscriptstyle {\rm S}}}^{\|}+2\partial_{(i}\chi_{{\scriptscriptstyle {\rm S}}j)}^{\bot}+\chi_{ij}^{\top}\right) &  \leftrightarrow  & \left(\Pi_{ij}+\frac{1}{2}G_{ij}^{tl}\right) \;.
\end{eqnarray*}
Note, however, that this correspondence is formal in the sense that the right--hand--side expressions are given in terms of local (Lagrangian) coordinates on the manifold that are constant along the solution, while the left--hand--side expressions are given in terms of global (Eulerian) background coordinates.

In order to better compare the standard results to ours, we modify the restrictions imposed
in \cite{Matarrese98}. First, we use a different gauge choice, which
is $\Delta\chi_{{\scriptscriptstyle {\rm S}}{\rm \bf i}}^{\|}=0$ instead
of $\Delta\chi_{{\scriptscriptstyle {\rm S}}{\rm \bf i}}^{\|}=-2\initial{\delta}$.
Second, we include the decaying mode. Then, the potential is no longer
independent of time and $\Delta\varphi({\bf x},t)=2 / (3\initial{t}^{2}a) \delta({\bf x},t)$.
Therefore, $\varphi$ now has two components $\varphi_{1}$ and $\varphi_{2}$
that are determined by $\Delta\varphi_{1}({\bf x})+a^{-5/2}\Delta\varphi_{2}({\bf x})=2 / (3 \initial{t}^{2}a)\delta({\bf x},t)$.
Finally, we include the prefactor on the right--hand--side into the definition
of a renormalized potential $\psi\left({\bf x},t\right)=\psi_{1}({\bf x})+a^{-5/2}\psi_{2}\left({\bf x}\right)$,
which then satisfies $\Delta\psi_{1}({\bf x})+a^{-5/2}\Delta\psi_{2}\left({\bf x}\right)=-a^{-1}\delta({\bf x},t)$.

With these changes, the resulting metric perturbations read:
\[
\mathcal{D}_{ij}\chi_{{\scriptscriptstyle {\rm S}}}^{\|}=2\left(\left(t / \initial{t}\right)^{\frac{2}{3}}-1\right)\mathcal{D}_{ij}\psi_{1}+2\left(\left(t / \initial{t}\right)^{-1}-1\right)\mathcal{D}_{ij}\psi_{2}
\]
and
\begin{eqnarray*}
\phi_{{\scriptscriptstyle {\rm S}}} &=& -10 / (9\initial{t}^{2})\psi_{1}
-\frac{1}{3}\left(\left(t / \initial{t}\right)^{\frac{2}{3}}-1\right)\Delta\psi_{1}\\
&&-\frac{1}{3}\left(\left(t / \initial{t}\right)^{-1}-1\right)\Delta\psi_{2}\;\;.
\end{eqnarray*}
The scalar and tensor fields are formally linked, in the comoving synchronous gauge, to the first--order metric and intrinsic perturbation field components of ${}^E \mathbf{P}^a$ and ${}^H \mathbf{\Pi}^a$, providing a dictionary between our formalism and the standard notation:
\begin{eqnarray}
\label{dictionary}
{}^E P_{ij} &  \leftrightarrow  &  \lbrack\left(t / \initial{t}\right)^{\frac{2}{3}}-1\rbrack \psi_{1\vert ij}+\lbrack\left(t / \initial{t}\right)^{-1}-1\rbrack\psi_{2\vert ij} \,;\nonumber\\
{}^H \Pi_{ij} &  \leftrightarrow  & \frac{1}{2}\left(\chi_{ij}^{\top}(t)-\chi_{ij}^{\top}(\initial{t})\right),
\end{eqnarray}
together with the correspondence with our initial data fields:
\begin{eqnarray}
\label{dictionaryinitial}
G_{ij} &  \leftrightarrow  & \frac{20}{9\initial{t}^{2}}\psi_{1}\delta_{ij}+ \chi_{ij}^{\top}(\initial{t})\nonumber\\
U_{ij} &  \leftrightarrow  & \frac{2}{3}\frac{1}{\initial{t}}\psi_{1\vert ij}-\frac{1}{\initial{t}}\psi_{2\vert ij}+\frac{1}{2}\dot{\chi}_{ij}^{\top}(\initial{t})\nonumber\\
W_{ij} &  \leftrightarrow  & \frac{2}{3}\frac{1}{\initial{t}^{2}}\lbrack\psi_{1\vert ij}+\psi_{2\vert ij}\rbrack+\frac{1}{2}\ddot{\chi}_{ij}^{\top}(\initial{t})+\initial{H}\dot{\chi}_{ij}^{\top}(\initial{t})\;.\nonumber\\\label{eq:dicW}
\end{eqnarray}
These three initial fields are not independent. Combining Equation~(\ref{kinematicRicci}) with Equation~(\ref{eq:riccifirst}) at $t=\initial{t}$, gives a differential equation that links $G_{ij}$ to $U_{ij}$ and $W_{ij}$. We have checked that this equation is consistent with the three relations (\ref{dictionaryinitial}). Taking the trace of the third of Equations~(\ref{eq:dicW}) and using $\Delta\psi_{1}({\bf x})+\Delta\psi_{2}\left({\bf x}\right)=-\delta({\bf x},\initial{t})$ provide another consistency check. We find $W^k_{\ k}\leftrightarrow - ( 3 / 2 ) \initial{H}^2 \delta({\bf x},\initial{t})$ (which is Equation~(49) in \cite{rza3}).

\subsection{Discussion of the difference between standard (Eulerian) and the intrinsic (Lagrangian) approaches}
\label{trafo}

Let us first illustrate the difference between a Lagrangian and an Eulerian approach in a flat--space situation. We take the example of 
inertial motion that, for a continuum, is governed by the following evolution equations for the three velocity field components $v^a$, indexed by the counter $a=1,2,3$ (a comma denotes partial derivative with respect to time $t$ or with respect to Eulerian coordinates $x^i$, respectively):
\begin{equation}
\label{inertial}
\frac{\mathrm d}{{\mathrm d}t} v^a = v^a_{\ ,t} + \frac{{\mathrm d}x^i}{{\mathrm d}t} v^a_{\ ,x^i} = 0 \;\;;\;\;
\frac{{\mathrm d}x^i}{{\mathrm d}t} = v^i \;\;.
\end{equation}
${\mathrm d} / {\mathrm d}t$ denotes the total or Lagrangian time--derivative operator.
The linearized version of this equation on the Euclidean background $v^a_{\ ,t} = 0$ is solved by a constant velocity field in Eulerian coordinates, $v^a (x^i , t) = v^a (x^i , t_i)$.
Introducing Lagrangian coordinates that follow the solution curves, ${\mathrm d}X^i / {\mathrm d}t = 0$, the total (Lagrangian) time--derivative reduces to a partial time--derivative at fixed Lagrangian coordinates, so that Equation (\ref{inertial}) reduces to a set of decoupled ordinary and linear differential equations, $v^a_{\ , t }(X^i ,t) =0$. This allows us to find the general solution of the coupled set of partial and nonlinear differential equations (\ref{inertial}) in Eulerian space, as long as the mapping between the Eulerian and the Lagrangian coordinates, provided by the Galilean solution curves $f^a : = X^a + v^a (X^i ,t_i) (t - t_i )$, is invertible:
\begin{eqnarray}
v^a (X^i , t) = v^a (X^ i , t_i ) \;;\qquad\quad\ \; \\ \Rightarrow \quad v^a (x^i , t) = v^a ( X^i = (f^a)^{-1} [ x^ i , t], t_i ) \;\;.\quad
\end{eqnarray}
We see in this example that the linearized Eulerian and the linear Lagrangian equations have the same algebraic form, but the solution of the latter includes Eulerian nonlinearities, here the convective term. 
Thus, in the close vicinity of the initial time, both solutions coincide, but the Lagrangian solution moves away from the Eulerian solution, and this motion is dictated by the solution itself. For the discussion in the next subsection we note that these deviations have an {\it recursive} structure \cite{buchert:varenna}:
\begin{equation}
v^a (t, x^i ) = v^a (t_i , X^i + (t-t_i) v^a (t_i , x^i = X^i +  \cdots )) \;.
\end{equation}
This example gives an intuition for the flat--space case. Working on nonflat spaces with Cartan coframes, the local coordinate representation of the coframes $\boldsymbol{\eta}^a (X^i ,t)$ exists as a projection on the local cotangent spaces with exact basis $\lbrace \bd X^i \rbrace$. Nonetheless, the diffeomorphism $f^a$ to the flat--space coordinates $x^i = f^{a=i}$, as used above, does not exist, since the coframes are not exact forms,
$\boldsymbol{\eta}^a \ne \bd f^a$. This is the reason why we cannot map the standard solution to an intrinsic solution by a coordinate transformation. 
Notwithstanding, we can find a solution of the intrinsic equation as we are going to outline in the following subsection.

\subsection{How to find relativistic Lagrangian first--order solutions for gravitational waves}

In order to find explicit first--order Lagrangian solutions of, e.g., the master equation for gravitational waves,
\eqref{mastermagnetic}, we can exploit the algebraic similarity of the standard linear equation and the Lagrangian master equation.
To determine the solution of \eqref{mastermagnetic} on a curved manifold we need to prescribe eigenfunctions other than flat--space Fourier modes. This is possible for constant--curvature spaces, but in general situations of an inhomogeneous curvature the solution procedure is not obvious.
In the cases where an equation on flat-space has algebraically the same form as an equation on a curved manifold, it is possible to find a solution. In what follows, we outline the procedure of how to find the 
general solution of the master equation \eqref{mastermagnetic},
$$
{}^H\ddot{\Pi}_{ij}+3H \ {}^H\dot{\Pi}_{ij} - a^{-2}\ {}^H{\Pi}_{ij} {}^{\vert k}_{\ \vert k} =a^{-2}\left({}^HW_{ij}^{tl} +\initial H {}^HU_{ij}^{tl}  \right).
$$
\begin{itemize}
\item
We employ the dictionary, i.e. the last equation of \eqref{dictionary} to redefine the dependent variable,
$$
{}^H \Pi_{ij} \;\; \leftrightarrow  \;\; \frac{1}{2}\left(\chi_{ij}^{\top}(\mathbf{x},t)-\chi_{ij}^{\top}(\mathbf{x},t_{i})\right).
$$
Equation~\eqref{mastermagnetic} is then cast into the standard linear propagation equation \eqref{eq:GWtEvol}, except that the former is an equation for functions of local (Lagrangian) coordinates,
and the latter is an equation for functions of (Eulerian) background coordinates.

\item
The linear propagation equation on flat space in the standard perturbation theory, Equation~\eqref{eq:GWtEvol}, can be solved. The solution in the Fourier space is provided by Equation~\eqref{eq:freegw-2}.
From this we have to prescribe initial data and perform the inverse Fourier transformation to obtain the solution as a function of Eulerian coordinates.

\item
The fact that our master equation \eqref{mastermagnetic} is algebraically identical (up to the redefinition of the dependent variable) to the Eulerian linear equation \eqref{eq:GWtEvol}, entitles us to take this Eulerian solution and replace the background coordinates by Lagrangian ones. 
The resulting solution is transported along geodesics, which are themselves solutions of this equation. They therefore exhibit a nonlinear behavior when seen from the background space reference frame.
In practice, we would prescribe the same initial data in the Lagrangian solution, since, initially, 
the Eulerian and Lagrangian coordinates coincide by definition. 
\end{itemize}

The above outlined solution procedure is not explicitly carried out in examples. We shall return to similar equations in the follow--up paper that includes pressure.

\section{Discussion and Outlook}
\label{sec:conclusion}

As perturbations of space propagate in the perturbed physical spacetime, the equations that describe their dynamics are intrinsically recursive. The standard linear solution for gravitational waves  does not
capture the nonlinear nature of these perturbations, see the criticisms advanced in 
\cite{pereira} that provide arguments why the standard linear solution is at odds with the physical phenomenon, see also \cite{ashtekar}. The common way out is to go to second-- and higher--order solutions within the standard approach \cite{Matarrese98,SAtensorperturbations,pereirahigherorder}.
However, as we discussed in the previous section, the intrinsic Lagrangian approach enjoys more generality, even if the equations
were linearized. This fact can be exploited to combine the simplicity of a linear solution with the nonlinear character of the physical problem at hand. Thanks to the present Lagrangian intrinsic approach we are in the situation to uncover nonlinear features of the gravitational waves. In other words, the linear intrinsic solution contains components that would be higher--order contributions in the standard linear approach.

In order to find out the wealth of information revealed by the Lagrangian approach, we now explore some avenues toward a more general solution than the one outlined in the last section.

\subsection*{Extrapolation of the solution into the nonperturbative regime}
\label{extrapolation}

We can certainly go beyond the linear solution by considering higher--order Lagrangian perturbations, as done in the standard Eulerian approach.
However, we can already employ the linear solution supplemented by extrapolation procedures in the spirit of the Newtonian Zel'dovich approximation (see \cite{rza1} for the discussion of the relativistic generalization of this extrapolation idea).

We now discuss different possibilities to understand nonperturbative properties of gravitational waves in terms of different extrapolation strategies. 

\subsubsection{Functional extrapolation}

The first strategy follows the line of thoughts in \cite{rza1} and \cite{rza2}. The idea is to insert the linear Lagrangian solution into 
exact functionals of the fields of interest \cite{rza1}, and we may then study their average properties \cite{rza2}. 
With this strategy we obtain nonperturbative expressions from a linear perturbation. Consider, as an example, the spatial metric. This is a bilinear form of the Cartan coframes and a linearization of the metric would truncate a quadratic term, thus, leading to a functional expression that is not the metric corresponding to the Cartan coframe that describes the deformation to a given order. The proposal is to keep this quadratic term, since the basic system of equations is written in terms of the coframes only and a consistent perturbation theory already determines the approximation for the coframes to a given order. Another example is the curvature that can be functionally defined in terms of coframes. Looking at the first-order deformation of the coframes, the corresponding curvature is the full functional expression, not the truncated one (see again \cite{rza2} for an explicit demonstration). 
(A related paper on averaging the gravitational wave tensor part is \cite{brown}.)

\subsubsection{Iterative extrapolation}

The second strategy notices that a strict linearization, pursued in this paper, neglects essential physical properties of 
gravitational waves propagating on a curved background. For example, this linearization replaces the Laplace--de Rham 
operator, Equation \eqref{laplacederham}, for the (curvature--generating, \textit{cf.} Subsection~\ref{curvatureconsequences}) trace--free gravitomagnetic part ${}^H\boldsymbol{\Pi}^a$ by a simple Laplacian in local coordinates:
\begin{eqnarray}
\left( \boldsymbol{\Delta}^{\rm dR}\ {}^H\boldsymbol{\Pi}^{a} \right)_i =  \left( \mathbf{\Delta}{}^H\boldsymbol{\Pi}^a  \right)_i  +  {}^H\Pi^a_{\ k} \: \mathcal{R}^{k}_{\ i} \quad\nonumber\\\approx -\Delta_{0}\,{}^H
\Pi^{a \;(1)}_{\ i}  \;\;,
\end{eqnarray}
where the first--order Ricci tensor \eqref{fullRicci} does not contribute at first order here. (At higher orders, the resulting curvature correction would couple the gravitomagnetic part to the trace and the gravitoelectric part of the perturbations.)  It is clear that a propagation of gravitational waves in a curved space will in general imply a nontrivial Laplace--de Rham operator. A possible route toward understanding of these nonlinear properties consists in 
keeping the Lagrangian time--evolution at first order in the linearized master equation \eqref{mastermagnetic}, but considering the full Laplace--de Rham operator of the Cartan coframes in the curved space section, i.e. by replacing the local Laplacian in \eqref{mastermagnetic} by the Laplace--de Rham operator. An iterative procedure to solve the resulting equation is suggested: we calculate the perturbed Cartan deformation for all parts according to the linearized scheme presented in this paper, then evaluate the Laplace--de Rham operator for this perturbed deformation (including the coupling to the trace and gravitoelectric parts), and iterate the equation. An intermediate step consists in just considering the `rough Laplacian', thus, avoiding the coupling to the trace and the trace--free gravitoelectric solutions due to the curvature correction. This intermediate step already renders the master equation for the perturbations \eqref{mastermagnetic} nonlinear in local coordinates due to the presence of the covariant spatial derivatives. Such an iteration may converge to a nonlinear solution after a sufficient number of steps (see, however, \cite{iterationGR}, and related iteration procedures for the Newtonian problem \cite{buchert06L,buchert06}).

\smallskip\noindent
{\small {\bf Acknowledgements:}
This work is part of a project that, in the final stage, has received funding from the European Research Council (ERC) under the European Union's Horizon 2020 research and innovation program (grant agreement ERC advanced Grant No. 740021--ARTHUS, PI: TB). A part of this project was funded by the National Science Centre, Poland, under Grant No. 2014/13/B/ST9/0084.
FAR acknowledges support by the {\'E}cole Doctorale PHAST Lyon. FAR and TB acknowledge student exchange support from the French--Bavarian Cooperation Center, BFHZ Munich \url{http://www.bayern-france.org/}. AW acknowledges support from the German research organization
DFG, Grants No.~WI 4501/1--1 and No.~WI 4501/2--1. \\
We thank Alexandre Alles for his helpful participation in the beginning stage of this paper.
Particular thanks go to L\'eo Brunswic, Mauro Carfora and Pierre Mourier for many valuable remarks and mathematical insights.
We thank them and also Martin Kerscher, Jan Ostrowski, Slava Mukhanov, Herbert Wagner and David Wiltshire for valuable discussions related to this work and comments.}

\appendix

\section{Propagation equations from the Maxwell--Weyl formalism}
\label{sec:appA}

In this appendix, we will derive Lagrangian propagation equations for the gravitoelectric and gravitomagnetic spatial parts of the Weyl tensor by employing the Maxwell--Weyl formalism \cite{ellis:cargese,wainwrightellis}. Related to this, the reader may also consult \cite{maartens:gravwaves,maartens:GEM,dunsby:gravwaves,ehlersbuchert:weyl}.

\subsection*{Gravitoelectric and gravitomagnetic parts of the Weyl tensor}

The Riemann curvature tensor can be split into a trace part, the Ricci tensor, that is linked to the matter content of the Universe by the Einstein field equations, and a traceless part, the Weyl tensor. This latter tensor, which encodes the relativistic counterpart of the Newtonian tidal forces and the gravitational waves, can be expressed through the Riemann curvature tensor as follows:
\begin{equation}
 C^{\mu \nu}_{\ \ \varrho\sigma} = \,\!^{(4)}\!R^{\mu \nu}_{\ \ \varrho\sigma} - 2\, \delta^{[\mu}_{\ [\varrho} \,\!^{(4)}\!R^{\nu]}_{\ \sigma]} +
\frac{1}{3} \delta^{[\mu}_{\ [\varrho} \delta^{\nu]}_{\ \sigma]} \,\!^{(4)}\!R \; .
\end{equation}
Let $u^{\mu}$ be the $4$--velocity of the fluid. The Weyl tensor can be split in an irreducible way into gravitoelectric and gravitomagnetic parts, $ E_{\mu \nu}$ and $ H_{\mu \nu}$, respectively, defined as follows: 
\begin{eqnarray}
 E_{\mu \nu} := C_{\mu \varrho \nu \sigma} \, u^{\varrho} u^{\sigma} \;;\\
 H_{\mu \nu} :=\, \frac{1}{2} \sqrt{| {}^{(4)} g |}\, \epsilon_{\alpha \beta \varrho (\mu } \, C^{\alpha \beta}_{\ \ \ \nu) \sigma}     \, u^{\varrho} u^{\sigma} \;,
\end{eqnarray}
where ${}^{(4)} g$ represents the determinant of the $4-$metric tensor ${}^{(4)} \mathbf{g}$, and $\epsilon_{\varrho \mu \alpha \beta} $ is the $4-$dimensional Levi--Civit\`a symbol.
As these parts are symmetric and traceless, they satisfy by construction the following identities:
\begin{equation*}
\begin{split}
 E^{\mu }_{\ \ \mu}  =  0 \:\:\:; \:\:\: E_{\mu \nu} = \ E_{(\mu \nu)}  \:\:\:; \:\:\: E_{\mu \nu}\ u^\mu =0\:\:\:,\\
 H^{\mu }_{\ \ \mu}  =  0\:\:\:; \:\:\:  H_{\mu \nu} = \ H_{(\mu \nu)} \:\:\:; \:\:\: H_{\mu \nu}\ u^\mu =0\:\:\:.
 \end{split}
\end{equation*}

\subsection*{Propagation equations for gravitoelectric and gravitomagnetic parts}

Combining Bianchi identities with the Einstein field equations we obtain:
\begin{equation}
\nabla^\kappa  C_{\mu \nu \kappa \lambda} = 8 \pi G \left( \nabla_{[\mu} T_{\nu] \lambda} +\frac13  g_{\lambda [\mu} \nabla_{\nu ]} T^{\kappa}_{\ \kappa} \right)\:\:.
\label{coucou}
\end{equation}
This equation can be turned into an equation for the gravitoelectric and gravitomagnetic parts, as the Weyl tensor can be expressed as follows:
\begin{eqnarray}
   \t{C}{-}{\mu\nu\kappa\lambda} & = & \left( \t{g}{-}{\mu\nu\alpha\beta} \t{g}{-}{\kappa\lambda\gamma\delta} - \t{\varepsilon}{-}{\mu\nu\alpha\beta} \t{\varepsilon}{-}{\kappa\lambda\gamma\delta} \right) \t{u}{\alpha}{} \t{u}{\gamma}{} \t{E}{{\beta\delta}}{} \\
   &+& \left( \t{\varepsilon}{-}{\mu\nu\alpha\beta} \t{g}{-}{\kappa\lambda\gamma\delta} + \t{g}{-}{\mu\nu\alpha\beta} \t{\varepsilon}{-}{\kappa\lambda\gamma\delta} \right) \t{u}{\alpha}{} \t{u}{\gamma}{} \t{H}{{\beta\delta}}{} \,,\; {\rm with}\nonumber
\end{eqnarray} 
$\t{g}{-}{\mu\nu\alpha\beta} \equiv \t{g}{-}{\mu\alpha} \t{g}{-}{\nu\beta} - \t{g}{-}{\mu\beta} \t{g}{-}{\nu\alpha}$, and
$\t{\varepsilon}{-}{\mu\nu\kappa\lambda} = \sqrt{-\mbox{}^{(4)}g}\ \t{\epsilon}{-}{\mu\nu\kappa\lambda}$.\\

In the rest frame of the dust fluid without vorticity, Equation (\ref{coucou}) yields the following constraints and evolution equations, e.g. \cite{bertschingerhamilton}:
\begin{eqnarray}
   \t{E}{k}{i \parallel k} - \t{g}{-}{ik} \t{\varepsilon}{kmn}{} \t{\Theta}{-}{ml} \t{H}{l}{n} &=& \frac{8 \pi G}{3} \frac{{\initial{\varrho}}_{\mid i}}{J}  \; ;\qquad \label{divE} \\
   \t{H}{k}{i \parallel k} + \t{g}{-}{ik} \t{\varepsilon}{kmn}{} \t{\Theta}{-}{ml} \t{E}{l}{n} &=& 0 \;\label{divH} \;;
\end{eqnarray}\\
\begin{eqnarray}
&\tdot{E}{-}{ij} + 2 \Theta \t{E}{-}{ij} - 3 \t{\Theta}{-}{k(i} \t{E}{k}{j)} - \t{\Theta}{k}{l} \t{E}{l}{k} \t{g}{-}{ij} \nonumber \\
  & = \t{g}{-}{m(i} \t{\varepsilon}{mkl}{} \t{H}{-}{j)l \parallel k} \label{dEdt} - 4 \pi G \frac{\initial{\varrho}}{J}  \Big( \t{\Theta}{-}{ij} - \frac{1}{3} \Theta \t{g}{-}{ij} \Big)\;,\qquad\quad \label{Edot} \\
&\tdot{H}{-}{ij} + 2 \Theta \t{H}{-}{ij} - 3 \t{\Theta}{-}{k(i} \t{H}{k}{j)} - \t{\Theta}{k}{l} \t{H}{l}{k} \t{g}{-}{ij}
 \nonumber\\  &= - \t{g}{-}{m(i} \t{\varepsilon}{mkl}{} \t{E}{-}{j)l \parallel k} \;. \label{Hdot}
\end{eqnarray}
Note that, in order to simplify (\ref{Hdot}), we used the identity 
$\initial{\varrho}_{\mid k}  \t{\varepsilon}{mkl}{} \t{g}{-}{m(i} \t{g}{-}{j)l} = 0$,
which stems from the antisymmetry of the Levi--Civit\`a symbol.

The expansion tensor can be split into its kinematical parts (with the shear $\sigma^i_{\; j}$ and the rate of expansion
$\Theta$, for vanishing vorticity):
\begin{equation}
\Theta^i_{\; j} = \sigma^i_{\; j}+\tfrac{1}{3} \Theta \delta^i_{\ j}\:\:.
\end{equation}
The initial density field $\initial{\varrho}$ can be split into $(\varrho_H)_{\rm\bf i} + \delta \initial{\varrho}$. At first order, the perturbation of the background density $\delta \initial{\varrho}$ can be expressed through
the trace of the nonintegrable generalization of the Newtonian acceleration gradient $\mathbf{W}^a$, \textit{cf.} Equation (\ref{initialP}), the generalization of the Poisson equation, $W = -4 \pi G \delta\initial{\varrho}$.

As ${E}_{ij}$ and ${H}_{ij}$ are symmetric and traceless tensors, at first order the constraints and evolution equations reduce to the following set: 
\begin{eqnarray}   
   \t{E}{k}{i \vert k}  &=&  - \tfrac{2}{3 a^3} \, W_{\:\vert i}\:;\label{divElin}   \\
      \t{H}{k}{i \vert k}   &=& 0 \:; \label{divHlin} \\
   \tdot{E}{-}{ij} +3 H {E}_{ij}  - \tfrac{1}{a} \epsilon_{(i}{}^{\,\,kl}H_{j)l \vert k}   &=& - 4 \pi G {\varrho}_H\, \sigma_{ij} \:; \quad\label{rotHlin} \\
   \tdot{H}{-}{ij} +3 H {H}_{ij}  + \tfrac{1}{a} \epsilon_{(i}{}^{\,\,kl} E_{j)l \vert k}   &=& 0\label{rotElin} \:,\\ \nonumber
\end{eqnarray}
where ${\varrho}_H =({\varrho}_H)_{\rm\bf i}\,/\,a^3 $ denotes the homogeneous part of the density field.
In order to obtain a propagation equation for the gravitoelectric spatial part of the Weyl tensor $E_{ij}$, we first multiply the  Equation (\ref{rotHlin}) by $a^4(t)$ and perform a time--derivative. Then, we inject Equation (\ref{rotElin}) and simplify the product of the Levi--Civit\`a symbols using the following identity:
\begin{equation}
\begin{split}
\epsilon_{kl (i}\epsilon_{j) mn}
=\delta_{ij}\delta_{km}\delta_{ln}-\delta_{ij}\delta_{lm}\delta_{kn}-\delta_{k(i}\delta_{j)m}\delta_{ln}\\
+\delta_{k(i}\delta_{j)n}\delta_{lm}+\delta_{l(i}\delta_{j)m}\delta_{kn}-\delta_{l(i}\delta_{j)n}\delta_{km} \;.
\end{split}
\end{equation}
Finally, using Equation (\ref{divElin}), we obtain the following second--order propagation equation: 
\begin{equation}
\begin{split}
\label{Efinal}
    \Box_0 E_{ij}  - 7 H \dot{E}_{ij} - 4 (4\pi G {\varrho}_H+ \Lambda) E_{ij} \\ 
    = - \frac{1}{a^5} \mathcal{D}_{ij} W  + 4 \pi G H {\varrho}_H \,\sigma_{ij} \,,
    \end{split}
\end{equation}
where $\Box_0 {X}_{ij} : = -\ddot{X}_{ij} + \tfrac{1}{a^2} \Delta_0 {X}_{ij}$ denotes the Lagrangian d'Alembertian applied to the tensor field ${X}_{ij}$.
To obtain the previous equation, we used the Friedmann equations to replace $3H^2 +  \ddot{a} / a$ by $20 \pi G {\varrho}_H / 3  + 4 \Lambda / 3$ (at order $0$ the curvature constant $k$ is assumed null). Furthermore, we used the compact notation $\mathcal{D}_{ij} W$ for $ W_{\vert ij}- 1/3 \delta_{ij} \Delta_0  W$.

The same logic applied to $H_{ij}$ leads to the following second--order propagation equation:
\begin{equation}
\begin{split}
\label{Hfinal}
 \Box_0 {H}_{ij} - 7 H \dot{H}_{ij} - 4 (5\pi G {\varrho}_H+ \Lambda) H_{ij} \\=  - \frac{1}{a}4 \pi G {\varrho}_H \epsilon_{(i}{}^{\: kl} \sigma_{j)l\vert k} \,.
    \end{split}
\end{equation}
(The coefficients in front of the third terms in Equations \eqref{Efinal} and \eqref{Hfinal} differ, since the first equation
features a term $\propto {\dot\sigma}_{ij}$ that can be replaced through ${\dot\sigma}_{ij} = - E_{ij}$, \textit{cf.} Equation (111) in \cite{rza1}, which changes the coefficient $5\pi G {\varrho}_H+ \Lambda$ to $4\pi G {\varrho}_H+ \Lambda$.)

\section{Erratum to \cite{rza3}}
\label{sec:appB}
In view of the definition $J \equiv \tfrac{\sqrt{g}}{\sqrt{G}}$ employed in this paper as well as in \cite{rza2,rza3}, the momentum constraint equations (27) of \cite{rza3} must contain factors of $\tfrac{1}{J}$, written correctly in Equation \eqref{form_momcoeff} (in \cite{rza1} the definition $J\equiv \sqrt{g}$ was used so that the inclusion of the factors is not needed).

\end{document}